\documentclass[preprint2]{aastex}
\def\({\left(}
\def\){\right)}
\def\[{\left[}
\def\]{\right]}
\begin{document}
\title {A simple model for magnetic reconnection heated corona}
\author{ B. F. Liu and S. Mineshige}
\affil{Yukawa Institute for Theoretical Physics, Kyoto University,
Kyoto 606-8502, Japan}
\email{bfliu@yukawa.kyoto-u.ac.jp}
\and 
 \author{K. Shibata}
\affil{Kwasan Observatory, Kyoto University, Yamashina, Kyoto 607-8471, Japan}
\begin{abstract}
We construct a simple model for  a magnetic reconnection heated
 corona above a thin 
accretion disk in AGNs 
and Galactic black hole candidates(GBHCs). The basic assumptions are
that (1) the magnetic reconnection heat is 
 cooled down overwhelmingly by Compton scattering in the
corona and that (2) thermal conduction is dominantly cooled  by
 evaporation of the chromospheric plasma in the disk-corona
 interface before Compton cooling sets in. 
With these two basic equations as well as equi-partition of magnetic
 energy with gas energy in the disk, 
 we can consistently determine the fraction of  accretion
energy dissipated in the corona without free parameters, and thus
 determine the temperature and 
 all other quantities in
both the corona and disk for given black hole mass and 
accretion rate. Then, we calculate the luminosity contributed   from
 the disk and 
 corona and the coronal flux weighted Compton $y$ 
parameter. It is found that, at a low luminosity  ($L<0.1L_{\rm Edd}$)
 the
 spectrum is hard with spectral index of
 $\alpha_{\rm sp}\sim 1$ ($f_\nu\propto \nu^{-\alpha_{\rm sp}}$),
 while  at a high
luminosity ($L\ga 0.1L_{\rm Edd}$) the spectrum can be either  soft or hard.
We also 
find that the situation is almost same for supermassive and stellar-mass black holes. 
 These features are consistent with observations
of AGNs and GBHCs. 
\end{abstract}

\section{Introduction}

There is a common belief that the hard X-ray radiation of GBHCs
 and  active galactic nuclei (AGNs) 
arises from hot gas around black holes,
either in forms of hot accretion flow or accretion-disk corona 
(e.g. Liang \& Nolan 1984; Mushotzsky, Done, \& Pounds 1993;
Narayan, Mahadevan, \& Quataert 1998).
The structure of hot accretion flow has been extensively investigated  
recently based on the vertically one-zone approximation.
Many calculations have so far been presented and various notions, 
such as ADAF, CDAF, and ADIOS, have been proposed 
(Ichimaru 1977; Blandford \& Begelman 1999;
Igumenshchev \& Abramowicz 2000).
Such studies can be directly compared with observations.  
The ADAF model, for example, can reasonably fit  the
observed spectra of low luminosity AGNs and 
X-ray novae in quiescence.
Nevertheless, co-existing soft and hard X-ray components
observed in ordinary AGNs and at least some GBHCs cannot
be well described by simple one-zone hot flow models and require
composite disk-corona structure.
Yet, we do not have a standard way of modeling disk corona
despite many attractive attempts
(e.g. Haardt \& Maraschi 1991; Meyer, Liu \& Meyer-Hofmeister 2000;
 R\'o\.za\'nska \& Czerny 2000; Kawaguchi, Shimura, \& Mineshige 2001).
One of the reasons is that, in addition to complex radiation
and energy interaction between the disk main body and corona,
magnetic fields, which tend to produce time variations
and spatial inhomogeneity (Kawaguchi et al. 2000),
seem to play essential roles there and are thus difficult to model
in a simple way.

There are good reasons to believe that accretion disk corona
may share some common features with solar corona 
(e.g., Galeev, Rosner, \& Vaiana 1979; Zhang et al. 2000). For example,
violent fluctuations are observed in both coronae (Takahara 1979) and
the fluctuation properties are remarkably similar in the
power-spectral
density plot (UeNo et al. 1997). 
Interesting theoretical investigations have also been presented 
in the frame of magnetized disk corona (e.g. Di Matteo 1998; Di
Matteo, Celotti, \& Fabian 1999; Miller \& Stone 2000; Merloni \& Fabian 2002).
Then,  
to what extent the accretion disk corona is similar to the solar
corona
in terms of basic physical processes?
What can we learn from the study of the solar corona for modeling
the accretion disk corona?

There has been a long history of the solar corona research, and recently
our knowledge on its structure and time variations
has been enormously enriched
thanks to the intensive X-ray observations from the space by 
the $Yohkoh$ satellite (Uchida, Kosugi, \& Hudson 1996).
To understand the observed emission measure and temperature relation
(EM-$T$ relation),
Shibata \& Yokoyama (1999) propose a magnetic reconnection 
model for solar flares.
They point out that only two equations suffice to explain the
observed EM-$T$ relation, which are the
energy balance between reconnection heating of 
magnetic field and conductive cooling, and
the energy equi-partition between gas and magnetic
fields.
They find the theoretical EM-$T$ relation in good agreement with
the observation of solar and stellar flares
over more than 10 orders of magnitudes in EM (see also Yokoyama \&
Shibata 2001). Furthermore, the coronal density can be estimated from the
balance between conductive heating and evaporation cooling of chromospheric plasma,
which is also consistent with observations.

With this  potentially powerful tool 
of describing the basic properties of the solar corona, we are tempted
to understand the  accretion disk corona along a similar line.
This is the starting point of this $Letter$.
Our aim is to propose a simple and self-consistent method, 
based on the interaction between the
disk and corona, to describe the basic properties of the disk and corona 
for given black hole mass and accretion rate,
with no free parameters constrained by observations. 
This is a big difference of our
attempt from previous magnetized disk corona models.  

\section{Magnetic reconnection heated corona}
First, we consider what relations determine the
thermal structure of magnetic reconnection heated corona.
From the X-ray observations of GBHCs and AGNs we know that 
the emitting temperature of hot
plasma is $\sim 10^9$K, and the power-law spectrum requires a Compton $y$
parameter of $y\equiv {4kT\over m_ec^2}\tau \sim 1$, leading to the 
 constraint on  the density of hot plasma to be 
$n\la 5\times 10^9 {\rm cm}^{-3}$  in a hot region 
of size $\ell\sim 10R_{\rm S}$ ($R_{\rm S}$ is Schwarzschild radius).  
A successful model should account for these observational data.

The energy balance in the magnetic flux tube holds
between heating by magnetic reconnection and
cooling by Compton scattering, thermal conduction, cycle-synchrotron
radiation. For a typical AGN system with  a black hole mass of $10^8
M_\odot$ and  an accretion rate of $0.1\dot M_{\rm
Edd}$ ($\dot M_{\rm Edd}=10L_{\rm Edd}/c^2$), 
the radiation energy density in disk is 
$U_{\rm rad}\approx 10^5{\rm ergs\  cm}^{-3}$.
The magnetic filed is $B\sim 10^3$G under the assumption of energy equi-partition with gas in the disk.
We find that the Compton scattering is the most efficient
cooling mechanism in the range of above parameters (which is also true in
GBHCs shown in Sect.4). We thus have 
\begin{equation}\label{e:energy}
{B^2\over 4\pi}V_A\approx {4kT\over m_ec^2}n\sigma_TcU_{\rm rad}\ell,
\end{equation} 
where $V_A$ is the Alfv\'en speed,
$U_{\rm rad}=U_{\rm rad}^{\rm in}+ U_{\rm rad}^{\rm re}$ the soft
photon filed to be Compton scattered from both the intrinsic disk
and reprocessed radiation, $\ell$ the
length of magnetic loop, and other constants have their standard meanings.
Therefore, the temperature in the magnetic flux tube is 
$T=1.21\times 10^9 n_9^{-{3/2}}U_5^{-1}B_3^3\ell_{14}^{-1}$K,
which is around the observed value of $\sim 10^9$K for typical
parameters of $n_9$,
$U_5$, $B_3$, and $\ell_{14}$ in units of $10^9{\rm cm}^{-3}$,
$ 10^5{\rm ergs\  cm}^{-3}$, $3\times 10^{14}{\rm cm} (=10R_{\rm S})$, respectively. 

Then how to determine the density? At the bottom of the flux tube,  structural calculations for 
both solar corona (Shmeleva \& Syrovatskii 1973)
and disk corona (Liu, Meyer, \& Meyer-Hofmeister 1995; Meyer et
al. 2000, Liu et al. 2002) show that there is a very thin transition
layer between the corona and the chromosphere, where temperature   
changes very steeply within a depth $h$ ($h\ll \ell$). 
The thermal conductive flux there is quite large. If the initial mass
density is too low, the conduction rate can exceed
the Compton cooling rate within this thin layer (i.e. conduction time
scale is less than Compton timescale). This conduction flux then
heats up some of the chromospheric plasma into the magnetic
tube 
in a similar way to the solar corona 
(Yokoyama \& Shibata 2001).  This process is called chromospheric
evaporation. The density of evaporated plasma  is
estimated from the energy balance at the interface, ${k_0T^{7/2}\over h}\approx
{\gamma\over \gamma-1} n_{\rm evap} k T {\(kT\over m_H\)}^{1/2}$ (where the evaporating speed is around sound speed due to the
large pressure gradient, $k_0\approx 10^{-6} {\rm
erg\,cm^{-1}\,s^{-1}\,K^{-7/2}}$, $\gamma=5/3$). This plasma eventually diffuses to the whole loop, and the
averaged density in the magnetic tube $n$ fulfills $n_{\rm
evap}h=n\ell$. We thus get,
\begin{equation}\label{e:evap}
{k_0T^{7\over 2}\over \ell}\approx {\gamma\over \gamma-1} n k T {\(kT\over m_H\)}^{1\over 2}.
\end{equation}
Once the density $n$ reaches a certain value, the Compton
cooling sets in and finally overwhelms the evaporation cooling, an
equilibrium is then established between the magnetic
heating and Compton cooling through the whole loop, evaporation at
the interface becomes quite slow or even stops. 
For a further check of the argument, we compare the enthalpy flux
with conductive flux from  numerical calculations on the vertical structure
of the frictionally heated  disk corona (Liu et al. 2002), 
finding that the approximation is correct in order of magnitudes.  
For $T=1.21
\times 10^9$K,  the density
$n$ given by Eq.(\ref{e:evap}) is $1.6\times 10^9{\rm cm}^{-3}$,
which  is just what is required. This implies
 that the evaporation can build up a
corona with a
density $\sim 10^9{\rm cm}^{-3}$.

Therefore, we expect two phases caused by the magnetic reconnection
heating: firstly, mass evaporation at the bottom of magnetic flux tube
quickly builds the corona up to a certain density; then, 
Compton scattering in the tube steadily radiates away the
magnetic heating.
Combining Eq.(\ref{e:energy}) with Eq.(\ref{e:evap}), we derive
$T$ and $n$ in the corona,
\begin{equation}\label{e:T}
T=1.02\times 10^9 U_5^{-{1\over
4}}B_3^{3\over 4}\ell_{14}^{1\over 8}K,
\end{equation} 
\begin{equation}\label{e:n}
n=1.12\times 10^9 U_5^{-{1\over
2}}B_3^{3\over 2}\ell_{14}^{-{3\over 4}} {\rm cm}^{-3}.
\end{equation} 
Eqs.(\ref{e:T}) and (\ref{e:n}) reproduce the typical temperature of 
$ 10^9$K and a density of $10^9 {\rm cm}^{-3}$ and
establish the relations
between the disk and the corona thorough disk radiation and magnetic
field. 

The magnetic energy is assumed to be equi-partitioned with gas energy
of the disk, $\beta\equiv n_{\rm disk} k T_{\rm disk}/{B^2\over 8\pi}=1$. 
Other possible values of $\beta$ have been studied by Miller \& Stone
(2000). In our model, larger $\beta$ means less energy dissipated
through the corona, and results in lower coronal $T$ and $n$ as
shown in Eqs.(\ref{e:Tr1})-(\ref{e:ng2}). To calculate disk quantities the
classical disk model (Shakura \& Sunyaev 1973) is used, in which
absorption is dominated by electron scattering and pressure by either
gas pressure $P_{\rm disk}^{\rm g}$ or radiation pressure $P_{\rm disk}^{\rm r}$, depending on the accretion rate.    
For $P_{\rm disk}=P_{\rm disk}^{\rm r}$, we find that the soft photon field is dominated by the intrinsic disk
radiation, and $T$ and $n$ in the corona are
\begin{equation}\label{e:Tr1}
T=1.36\times 10^9 \alpha_{0.3}^{-{15\over 32}}\beta_1^{-{3\over 8}}
m_8^{-{3\over 32}}\dot m_{0.1}^{-1}r_{10}^{75\over 64}\ell_{10}^{1\over 8}K
\end{equation}  
and
\begin{equation}\label{e:nr1}
n=2.01\times 10^9  \alpha_{0.3}^{-{15\over 16}}\beta_1^{-{3\over 4}}
m_8^{-{19\over 16}}\dot m_{0.1}^{-2}r_{10}^{75\over 32}\ell_{10}^{-{3\over 4}} {\rm cm}^{-3}
\end{equation} 
with standard parameters, viscous coefficient $\alpha_{0.3}$, $\beta_1, $$m_8$, $\dot m_{0.1}$, $r_{10}$, $\ell_{10}$ in
units of 0.3, 1, $10^8M_\odot$, $0.1\dot M_{\rm Edd}$,$10R_{\rm S}$,
$10R_{\rm S}$, respectively. For $P_{\rm disk}=P_{\rm disk}^{\rm g}$,  
the soft photon field is dominated by
the reprocessed coronal radiations $U_{\rm rad}\approx 0.4U_{\rm B}$
($\sim 40\%$ of the coronal radiation is reprocessed, see Haardt \&
				   Maraschi 1991)
and we have
\begin{equation}\label{e:Tg2}
T=4.86 \times 10^9  \alpha_{0.3}^{-{9\over 80}}\beta_1^{-{1\over 8}}
m_8^{{1\over 80}}\dot m_{0.1}^{1\over 10}r_{10}^{-{51\over 160}}\ell_{10}^{1\over 8}K,
\end{equation}
\begin{equation}\label{e:ng2}
n=2.55\times 10^{10} \alpha_{0.3}^{-{9\over 40}}\beta_1^{-{1\over 4}}
m_8^{-{39\over 40}}\dot m_{0.1}^{1\over 5}r_{10}^{-{51\over 80}}\ell_{10}^{-{3\over 4}} {\rm cm}^{-3}.
\end{equation} 

\section{Fraction of accretion energy released through corona}
So far, we have not considered a back reaction, that is, the energy transfered from the disk to the corona by magnetic
reconnection essentially affects the disk structure. 
Defining $f$ as the fraction of accretion energy released
in the reconnected magnetic corona, 
\begin{equation}\label{e:f-d}
f\equiv {F_{\rm cor}\over  F_{\rm tot}}=\({B^2\over 4\pi}V_A\) \({3GM\dot M\over
8\pi R^3}\)^{-1},
\end{equation}
we find that, in the expression of disk quantities, such as
pressure (and hence $B$) and $U_{\rm rad}$, $\dot m_{0.1} $ should be
replaced by $(1-f)\dot m_{0.1}$.
 Then the coronal
temperature and density in Eqs.(\ref{e:Tr1})-(\ref{e:ng2}),  should be
modified appropriately, which, in turn, affects the
amount of reprocessed radiation. Finally,
 we obtain an equation concerning $f$ from
Eq.(\ref{e:f-d}) for
a radiation pressure-dominant disk
\begin{equation}\label{e:fr1}
f=1.45(1-f)^{-2}\alpha_{0.3}^{-{45\over 32}}\beta_1^{-{9\over 8}}
m_8^{-{9\over 32}}\dot m_{0.1}^{{-3}}r_{10}^{{225\over
64}}\ell_{10}^{3\over 8},
\end{equation}
and for a gas pressure-dominant disk
\begin{equation}\label{e:fg}
f=4.70\times 10^4(1-f)^{11\over 10}
\alpha_{0.3}^{-{99\over 80}}\beta_1^{-{11\over 8}}
m_8^{{11\over 80}}\dot m_{0.1}^{{1\over 10}}r_{10}^{-{81\over
160}}\ell_{10}^{3\over 8}.
\end{equation}  
Therefore, the ratio $f$ is not a free parameter in our model and 
can be calculated out by solving Eq.(\ref{e:fr1}) or
Eq.(\ref{e:fg}) for given 
black hole mass and accretion rate.

With $0< f< 1$,  Eq.(\ref{e:fr1})
has either double solutions or no
solution.  Only at  small distances, $R<R_{\rm rad}$, there exist
double solutions, where
\begin{equation}\label{e:R-radin}
R_{\rm rad}=8R_{\rm S} \alpha_{0.3}^{{2\over 5}}\beta_1^{{8\over 25}}
m_8^{{2\over 25}}\dot m_{0.1}^{{64\over 75}}\ell_{10}^{-{8\over 75}}. 
\end{equation}
This is to say,  the radiation pressure-dominant disk 
exists only in the innermost region at relatively high $\dot m$. At
low  $\dot m$, the critical radius [Eq.(\ref{e:R-radin})] becomes less
than $3R_{\rm S}$, the radiation pressure-dominant region would no
 longer exist. 
In contrast, Eq.(\ref{e:fg}) always has unique solution, 
 indicating that the gas pressure-dominant region can
extend to very small radius even at high $\dot m$ as long as a large
fraction of accretion energy dissipates in the corona ($f\sim 1$). 

Fig.\ref{f:tny-r} shows the radial
distribution of variables $T$, $n$, $B$, $V_{\rm A}$,
and $\tau$ for $\dot M=0.1\dot M_{\rm Edd}$ (left)
and  $\dot M=0.5 \dot M_{\rm Edd}$ (right). At $0.1\dot M_{\rm Edd}$ there is no
radiation pressure-dominant disk even close to the last stable
orbit, and almost all the energy is transfered to the corona and radiated
away by Compton scattering. The temperature in the corona is
$\sim 2\times 10^9$K, the density  $\sim 3\times 10^9 {\rm cm}^{-3}$,
the magnetic field $\sim 10^3$G, optical depth $\sim 0.8$, and Alfv\'en
speed $\sim 0.2c$. The magnetic pressure before reconnection is
$\sim30$ gas pressure of the reconnected corona, which indicates a
$\sim30$ times vertical decrease of  gas pressure from the disk midplane to the
corona under the equi-partition principle.  
With increase of $\dot M$, all these quantities
increase slightly until the radiation pressure-dominant region appears in the
innermost region. Temperature and other quantities in this state are
lower than that at gas pressure-dominant state as shown in right panel of Fig.1.      

\section{Compton $y$ parameter and observational implications}
\begin{figure}
\plottwo{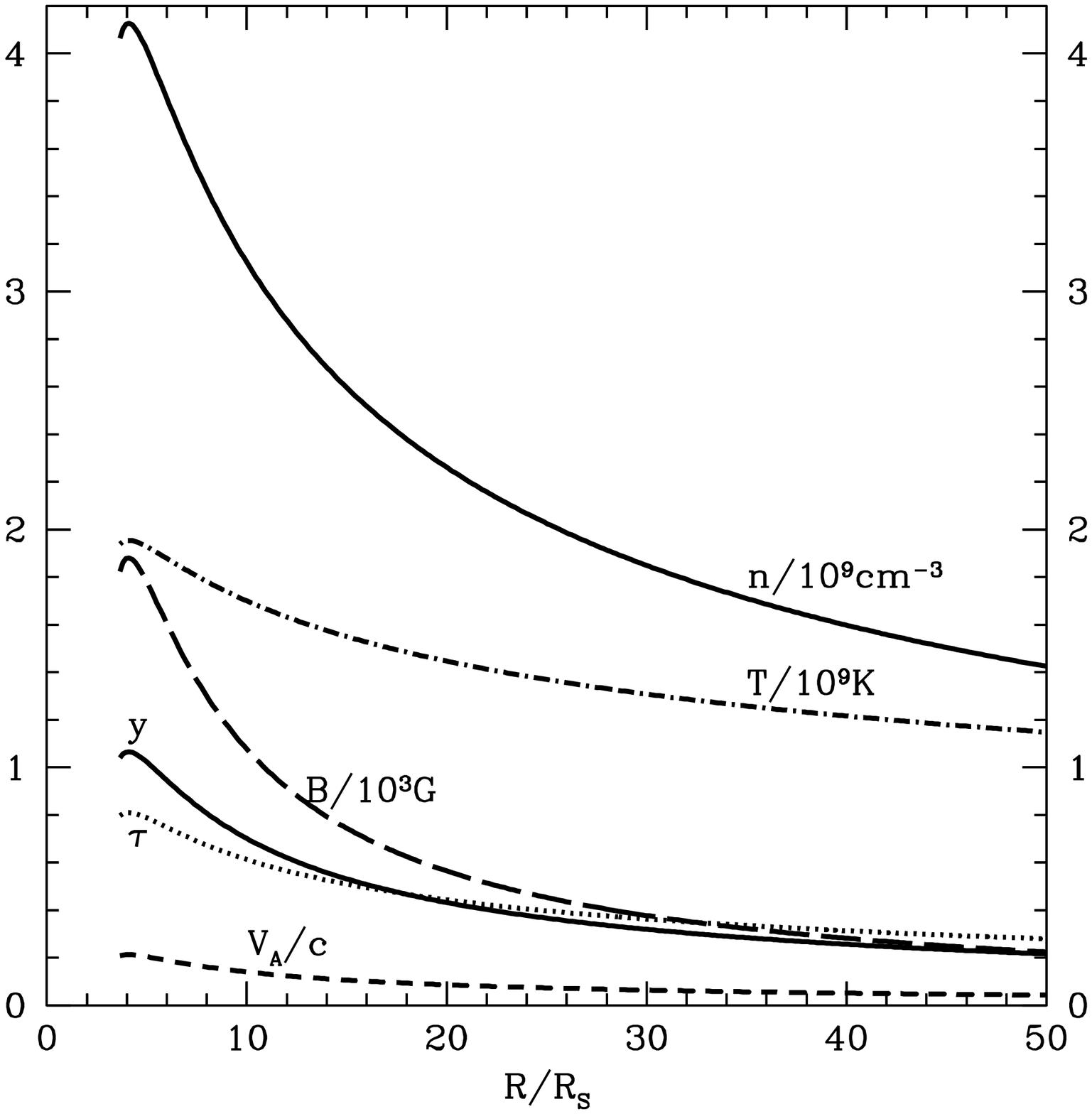}{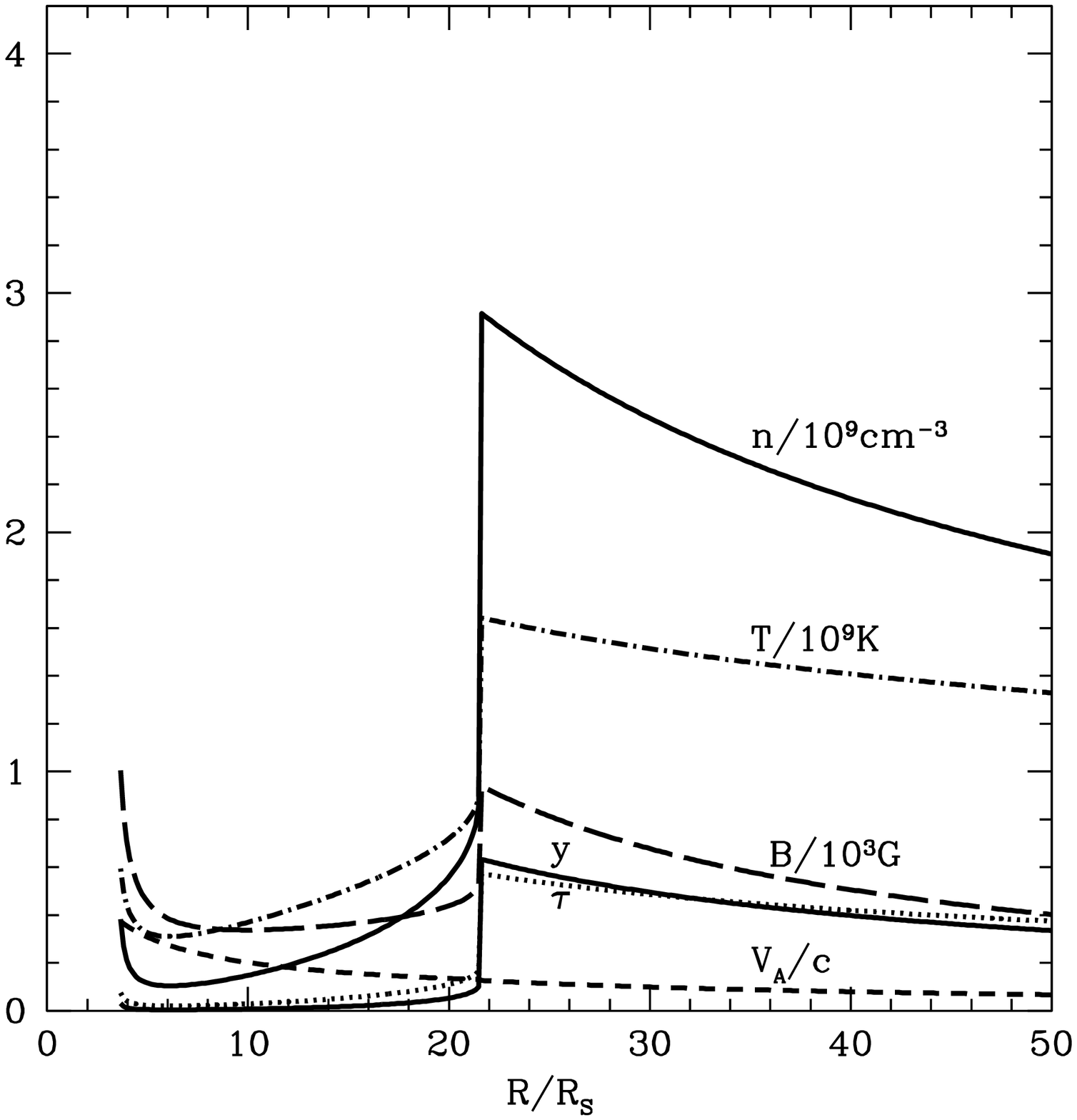}
\caption[]{\label{f:tny-r} Coronal quantities along distance from a
$10^8M_\odot$ black hole at accretion rate of $0.1\dot M_{\rm
Edd}$($=L_{\rm Edd}/c$, left panel)
and $0.5\dot M_{\rm
Edd}$  (right panel).}
\end{figure}

\begin{figure}
\plotone{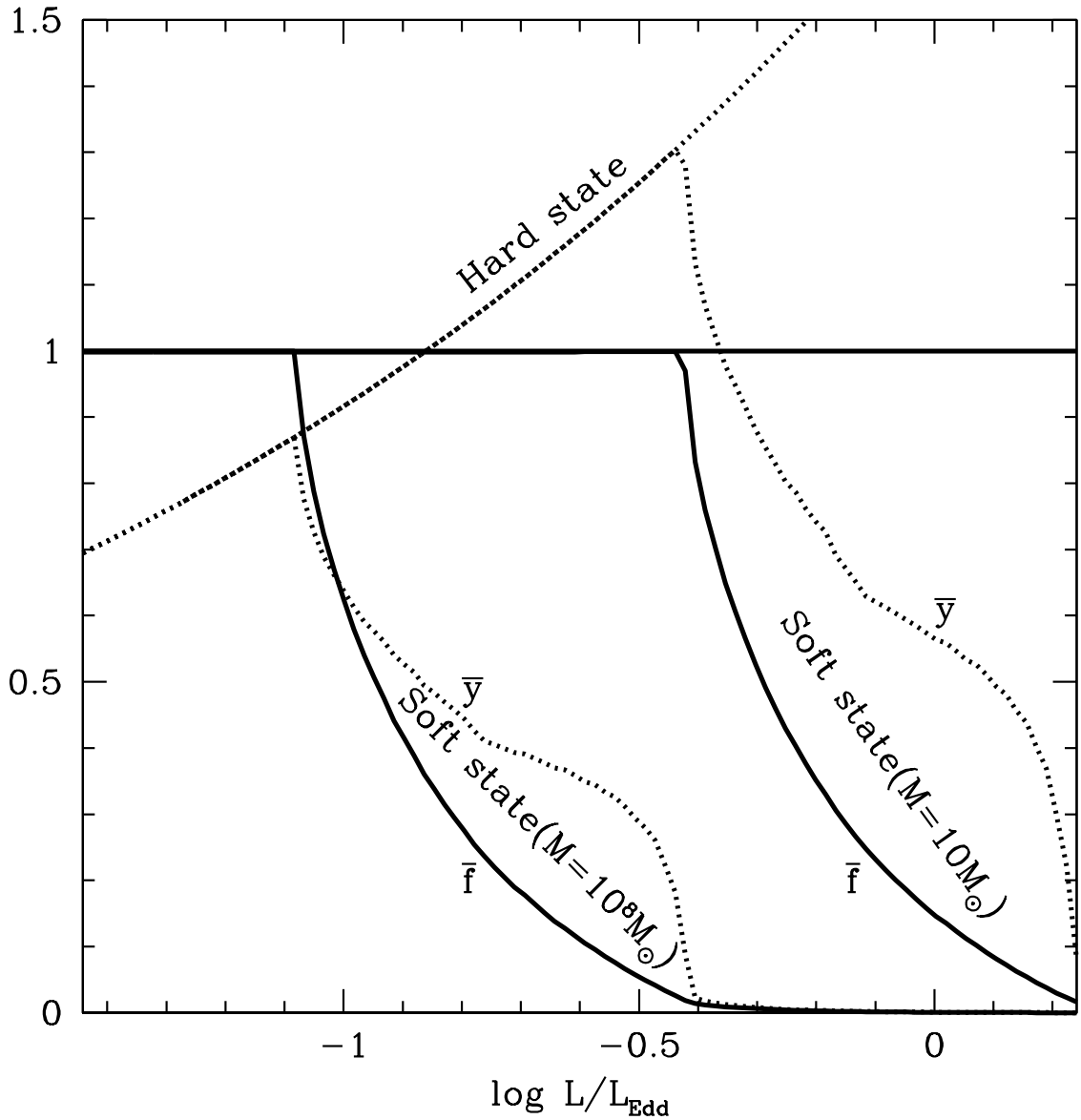}
\caption[]{\label{f:y-mdot} The fraction of total luminosity
released in the corona (solid curves) and the flux weighted Compton
$y$ parameter $\bar{y}$ (dotted curves) in
dependence on the mass accretion rate for  
a $10^8M_\odot$ black hole and a $10M_\odot$ black hole. Remarkably,
there is no difference between the cases with different black hole
masses at $L\la 0.1 L_{\rm Edd}$.}
\end{figure}

 The shape of the Compton spectrum depends
on the $y$ parameter. In an optically thin corona with 
thermal-distribution non-relativistic electrons, we have
$y={4kT\over m_e c^2} n \sigma_T\ell$ and spectral index $\alpha_{\rm sp}=\sqrt{{9\over 4}+{4\over y}}-{3\over 2}$.
Fig.\ref{f:tny-r} shows that the Compton $y$ parameter slowly
decreases outwards along distance. In order to compare the spectrum
with observations we calculate the coronal flux weighted
$\bar{y}$ parameter from $3R_{\rm S}$  to $50R_{\rm S}$ for a
series of  mass accretion rates.
Mean fraction 
$\bar{f}$  (weighted by total accretion energy) of accretion
energy radiating from the corona 
is also calculated so that we can know the relative
strength of luminosity
from disk and corona. 
Results plotted in Fig.\ref{f:y-mdot} show that 
$\bar{y}\sim$0.7 at $\dot M=0.1\dot M_{\rm
Edd}$, and that the accretion energy is overwhelmingly
dissipated in the corona ($\bar{f}\approx 1$) with a coronal
luminosity $L_{\rm cor}\approx 4.3\times 10^{44}{\rm ergs}\,{\rm
s}^{-1}$. At a rate of 0.5$\dot M_{\rm Edd}$, radiation
pressure-dominant region appears in the innermost part, where
$\bar{f}\approx 0.23$,
 $\bar{y}\approx 0.41$, and $L_{\rm cor}\approx 5.1\times 10^{44}{\rm ergs}\,{\rm s}^{-1}$.
At an even higher accretion rate, $\dot M=\dot M_{\rm
Edd}$, the radiation pressure-dominant part expands   and the emission of
the system is
dominated by the disk. We find $\bar{f}\approx 0.03$ and the coronal luminosity
is mainly contributed from outer region ($40R_{\rm S}$ to $50R_{\rm S}$) ,  
$L_{\rm cor}\approx 1.5\times 10^{44}{\rm ergs}\,{\rm s}^{-1}$, 
which is smaller than  that at $\dot M=0.1\dot M_{\rm Edd}$, and the corresponding   
$\bar{y}$($\approx 0.23$) is lower. 
The results imply that, for a high rate
the emergent spectrum  is soft: a large fraction of the accretion energy is
dissipated in the disk, the system is bright 
in optical-UV and soft X-rays contributed by the strong disk, 
but is faint in hard X-rays contributed by the weak corona, and thus
the spectrum is steep. While for a
low mass accretion rate, the spectrum is hard: almost all the accretion energy
is channeled to the magnetized corona and is radiated by Compton
scattering, the $\bar{y}$ parameter is 0.7--1, indicating a
hard X-ray tail. These features are summarized in Table \ref{t:state}.

We also show in Fig.\ref{f:y-mdot} that  
$\bar{f}$  and
$\bar{y}$ of the stellar-mass black hole  are the same as those of
supermassive black hole at low accretion rates, though the 
radiation pressure-dominant part appears  at a rate $\sim 3$ times larger than that of AGN.  

\begin{table} [ht]
\begin{center}
\caption[]{\label{t:state} Two types of solution}
\begin{tabular}{ccccc}
\noalign{\smallskip}\hline\hline\noalign{\smallskip}
Spectral&$\bar{f}$&Range  & Pressure&$\alpha_{\rm sp}$\\
state&& of $\dot M$&in disk&\\
\noalign{\smallskip}\hline\noalign{\smallskip}
Hard&$\sim 1$&Always&$P_{\rm disk}^{\rm g}$&$\sim 1$\\
Soft&$<0.5$& $>0.1\dot M_{\rm Edd}$&$P_{\rm disk}^{\rm r}$&$>2$\\
\noalign{\smallskip}\hline\hline\noalign{\smallskip} 
\end{tabular}
\end{center}
\end{table}
As we pointed out in Sect.3, there always exists gas pressure-dominant solution 
 with $T_{\rm disk}\sim 10^4K$ down to the last stable orbit
regardless of  accretion rates, as
 long as  a very large fraction of energy is channeled to the corona
(Similar result is obtained in other way by  Nakamura \& Osaki 1993
but for an assumed $f$ value),
and radiation pressure-dominant
solutions are found only at high accretion rate with a large fraction of
accretion energy dissipated in the disk.
Therefore, there are two solutions in the inner region if the mass
accretion rate is high enough: large $f$ corresponds to the hard spectrum
and small $f$ the soft spectrum.
 This might explain why some soft X-ray transients (SXTs),
such as GS 2023+338, continuously stay in the hard state throughout the
outburst, while in other SXTs the spectral transition occurs from soft to hard
during the decay and/or from hard to
soft during the rise of an outburst (Tanaka \& Shibazaki 1996).
Observations in AGNs also show weaker contributions to the hard X-rays for
objects with higher accretion rate like narrow-line Seyfert 1 galaxies
 which are thought to have large $\dot M/M$ (Boller 2000
and references therein).
 We thus
propose that the way of spectral state transition is not unique:
a) at very high luminosity ($L\sim L_{\rm Edd}$), both the corona and disk
are strong, which might be time-dependent; b) at moderately high luminosity 
($L\ga 0.1L_{\rm Edd}$),
the system is usually dominated by disk, for which the spectrum is soft; but
hard state is also possible, in which only a little accretion energy is
left in the disk (this is different from the proposition by Narayan
et al. (1998) who claim that  only the soft state is realized); c)at 
low luminosity ($L< 0.1L_{\rm Edd}$), most of the accretion energy is
transfered to the corona and the power-law type hard X-ray spectrum is produced.        

\acknowledgements{We thank Prof. R.Narayan for encouraging discussion. BFL would like to thank the Japan Society for the
Promotion of Science for support. This work is partially supported by
the Grants-in Aid of the Ministry of Education, Science, Sports, and
Culture of Japan (P.01020, BFL; 13640238, SM). }

\onecolumn
\clearpage 

\clearpage

\end{document}